\documentstyle[12pt,aaspp4]{article}

\def\Q{R_{M,e}}

\def\part{\partial}

\def\curl{\nabla \times}
\def\bfE{{\bf E}}
\def\bfJ{{\bf J}}

\def\bfB{{\bf B}}

\def\bV{\overline V}

\def\beqn{\begin{eqnarray}}
\def\eeqn{\end{eqnarray}}

\def\ni{\noindent}
\def\.{\mathaccent 95}
\def\beq{\begin{equation}}
\def\ee{\end{equation}}

\def\ta{\tau}

\def\Phi{\Phi}

\def\frac#1#2{{\textstyle{{#1}\over {#2}}}}
\def\ni{\noindent}
\def\lsim{\mathrel{\rlap{\lower4pt\hbox{\hskip1pt$\sim$}}
    \raise1pt\hbox{$<$}}}
\def\gsim{\mathrel{\rlap{\lower4pt\hbox{\hskip1pt$\sim$}}
    \raise1pt\hbox{$>$}}}
\def\sqr#1#2{{\vcenter{\vbox{\hrule height.#2pt
         \hbox{\vrule width.#2pt height#1pt \kern#1pt
         \vrule width.#2pt}
         \hrule height.#2pt}}}}

\newbox\grsign \setbox\grsign=\hbox{$>$} \newdimen\grdimen \grdimen=\ht\grsign
\newbox\simlessbox \newbox\simgreatbox
\setbox\simgreatbox=\hbox{\raise.5ex\hbox{$>$}\llap
     {\lower.5ex\hbox{$\sim$}}}\ht1=\grdimen\dp1=0pt
\setbox\simlessbox=\hbox{\raise.5ex\hbox{$<$}\llap
     {\lower.5ex\hbox{$\sim$}}}\ht2=\grdimen\dp2=0pt

\def\doublespace {\smallskipamount=6pt plus2pt minus2pt
                  \medskipamount=12pt plus4pt minus4pt
                  \bigskipamount=24pt plus8pt minus8pt
                  \normalbaselineskip=24pt plus0pt minus0pt
                  \normallineskip=2pt
                  \normallineskiplimit=0pt
                  \jot=6pt
                  {\def\smallskip {\vskip\smallskipamount}}
                  {\def\medskip   {\vskip\medskipamount}}
                  {\def\bigskip   {\vskip\bigskipamount}}
                  {\setbox\strutbox=\hbox{\vrule 
                    height17.0pt depth7.0pt width 0pt}}
                  \parskip 12.0pt
                  \normalbaselines}

\font\gkvec=cmmib10                         
\def\bomega{\hbox{{\gkvec\char33}}}                  

\def\lb{\langle}
\def\rb{\rangle}

\def\bw{\bar{\omega}}
\def\bv{\bar V}
\def\bB{\overline B}
\def\ts{\times}
\def\lb{\langle}
\def\rb{\rangle}
\def\curl{\nabla {\ts}}

\def\bbV{\overline {\bf V}}
\def\bfv{{\bf v}}

\def\bfj{{\bf j}}

\def\bfw{{\bomega}}

\def\bfb{{\bf b}}

\def\bfB{{\bf B}}

\def\bbB{\overline {\bf B}}
\def\nb{\nabla}
\def\curl{\nb\ts}

\def\b0{b^{(0)}}
\def\v0{v^{(0)}}
\def\w0{\omega^{(0)}}
\def\bb0{\bfb^{(0)}}
\def\bv0{\bfv^{(0)}}
\def\bw0{\bfw^{(0)}}
\def\bj0{\bfj^{(0)}}

\def\ni{\noindent}

\eject

\begin{document}

\title{Dimensionless Measures of Turbulent Magnetohydrodynamic Dissipation Rates}
\medskip
\author{Eric G. Blackman$^{1}$ and George B. Field$^2$}
\affil{1. Department of Physics \& Astronomy and Laboratory for Laser
Energetics, University of Rochester, Rochester NY 14627, USA;
2. Center for Astrophysics, 60 Garden St., Cambridge MA, 02139, USA}

(submitted to MNRAS)

\begin{abstract}
The  magnetic Reynolds number $R_M$, is defined as the product of a characteristic  scale and associated flow  speed  divided by the microphysical 
magnetic diffusivity. 
For laminar flows, $R_M$ also  approximates the ratio of advective to dissipative terms in the total  magnetic energy equation, but  for  turbulent flows this latter ratio  depends on the energy spectra and approaches unity in a steady state.  
{To generalize for flows of arbitrary spectra we define an effective magnetic dissipation number, $R_{M,e}$, as the ratio of the advection to microphysical dissipation terms  
in  the total magnetic energy equation, incorporating the full spectrum of scales, arbitrary magnetic Prandtl numbers,  and  distinct pairs of inner and outer scales for   magnetic and kinetic spectra.  As expected,  for a 
  substantial  parameter range $R_{M,e}\sim {O}(1) << R_M$.
We also distinguish  $R_{M,e}$ from ${\tilde R}_{M,e}$ where the latter 
is an effective magnetic Reynolds number  for the mean magnetic 
field equation when a turbulent diffusivity is explicitly imposed as a closure.
That  $R_{M,e}$ and ${\tilde R}_{M,e} $  approach unity even if
$R_M>>1$ highlights that, just as in hydrodynamic turbulence,
energy dissipation  of large scale structures 
 in turbulent flows via a cascade 
can be  much faster than the dissipation of large scale 
structures in laminar flows.  
 This illustrates that the 
rate of energy dissipation by magnetic reconnection
is much faster in turbulent flows, and much less sensitive 
to microphysical reconnection rates 
compared to laminar flows.}

\end{abstract}


\medskip


\bigskip

\section{Introduction}

Magnetic fields play an observable and dynamical role in a
range of astrophysical and laboratory plasmas.  When the scales of
interest are larger than the particle mean free path, magnetohydrodynamics
provides a suitable approximation (\cite{moffatt}, \cite{p79}, \cite{biskamp97}). 
Derived from Maxwell's equations and Ohm's law, the 
induction equation describing the time evolution of the magnetic field $\bfB$
in non-relativistic magnetohydrodynamics (MHD) is given by 
\beq
\partial_t\bfB= \curl(\bfv\ts\bfB) +\lambda\nabla^2\bfB,
\label{0}
\ee
where $\lambda$ is the magnetic diffusivity,
and $\bfv$ is the velocity, determined from the fluid momentum
equation. For incompressible flows, the momentum equation incorporating
the 
Lorentz force is
\beq
\partial_t\bfv=-\bfv\cdot\nabla \bfv-\nabla P/\rho+ (\curl\bfB)\ts\bfB/4\pi\rho +\nu \nabla^2\bfv,
\label{01}
\ee
where $P$ is the pressure and $\nu$ is the kinematic viscosity, and $\rho$ is the density.

Taking the approximate  ratio of magnitudes of the second to third terms
in (\ref{0})  introduces the magnetic Reynolds number
\beq
R_{M,l}\equiv l v/\lambda,
\label{oldrm}
\ee
where  $l$ is a macroscopic scale of field variation
for the system being studied. When $l$ is the largest
macroscopic scale of the system,  we write $R_M$, eliminating the subscript $l$.

For laminar flows in which $R_M>>1$,  the last term
of (\ref{0}) is often ignored.  An important consequence of ignoring this 
term is magnetic flux freezing, $d_t\int\bfB\cdot d{\bf S}=0$ (\cite{p79}) 
which implies that the integrated product of the magnetic field and
the surface area it penetrates, or the magnetic flux moving with the plasma,
is  conserved.  
Dotting (\ref{0}) with $\bfB$ gives the same $R_M$ for the corresponding
terms in the magnetic energy equation so hat  $R_M>>1$ 
would also imply that the resistive dissipation term is small compared to the other dynamical terms. When most of the magnetic energy is on the largest scale, the implication is that the dissipation of the total magnetic energy is negligible.

The above physical implications of  $R_M>>1$  do
not  apply if the flow is turbulent. 
When an initially laminar flow becomes turbulent, 
 the field and flow evolve 
to develop non-trivial turbulent spectra (e.g. \cite{biskamp97,biskamp03}). 
Eqn. (\ref{oldrm}), with $l$ taken as the largest
macroscopic scale of the system,   
is no longer a good approximation to the ratio of the advection
to dissipation term in the total magnetic energy equation even if the energy is dominated by structures on that scale. The 
 dissipation term is  not necessarily  negligible because it depends more strongly on high wavenumber components of the turbulent spectrum than do the
advection terms.

The importance of  dissipation in MHD turbulence just described, 
is directly analogous to the case of  hydrodynamic turbulence.
For the latter, the Kolmogorov steady-state assumption (Kolmogorov 1941;
see also Davidson 2004)
that the energy dissipation rate per mass is constant on all
scales through the cascade implies that the energy dissipation rate 
at the dissipation scales exactly equals that at the input scale. 
Therefore, the dissipation term in the total energy evolution equation
exactly balances the nonlinear energy transfer terms when all scales
are included and a steady-state is assumed. 
Extracting dimensionless measures of hydrodynamic turbulence dissipation rates
have confirmed this basic conceptual picture (e.g. Pearson et al. 2004).

In making contact with these concepts 
from hydrodynamic turbulence,  
we find it particularly important to emphasize that dissipation 
is non-negligible in MHD turbulence even though $R_M>>1$. This very much 
contrasts the physical implications commonly attributed to 
laminar $R_M>>1$  MHD flows.  
Characterizing the MHD energy depletion time scales
for large $R_M$ turbulent MHD flows such
as those of stellar convection zones, accretion disks, and galactic interstellar media, is important for understanding the  magnetic field evolution, conversion of magnetic energy into particles, and requirements for magnetic field sustenance.
Since these flows are characterized by a spectrum of scales (\cite{biskamp97}, \cite{biskamp03}), 
the ratio of advection to dissipation terms in the total magnetic energy equation  is not  in general $R_M$ 
but  a generalized magnetic dissipation  number $R_{M,e}$ 
which depends on the magnetic and kinetic energy spectra. 

Quantifying 
$R_{M,e}$,
requires incorporating the spectrum of magnetic and 
velocity fields along with  the magnetic Prandtl number $Pr_M\equiv\nu/\lambda$. 
{Here we calculate $R_{M,e}$ 
as a function of 
the magnetic spectral index and $Pr_M$, 
the latter of which depends on the microphysical diffusivities and thus on $R_M$.
Although $R_M$ (as defined with the largest scale) 
does help determine  whether the flow would have become turbulent in the first place, it is not a good indicator of the 
time scale  for overall  energy dissipation when the flow is turbulent.  
For regimes in which  $R_{M,e}$ is largely independent of $R_M$, 
the total energy depletion rate  becomes largely independent of 
the specific small scale mechanism of dissipation; 
turbulence cascades the energy
to scales where the dissipation time is short.
}

In section 2, we derive the approximations for the corresponding 
second and third terms of (\ref{0}) in the magnetic energy equation, incorporating the turbulent spectra 
and triple correlations for the case
without a mean field. We then plot $R_{M,e}$
as a function of the magnetic spectral index and $Pr_M$. 
In section 3, we  distinguish $R_{M,e}$  from a different  quantity
${\tilde R}_{M,e}$ for the mean field equation.
In section 4, we conclude with  a discussion of the implications
of ${\tilde R}_{M,e}<<1$ and $R_{M,e} <<R_M$.

\section{$R_{M,e}$ in the absence
of 
a mean field}


We write the magnetic field as a sum of  mean and fluctuating contributions
such that 
 $\bfB=\bbB+\bfb$ where $\bbB$ is the mean field
that survives the averaging over scales much larger than 
the scale of  any fluctuations incorporated in $\bfb$. 
We use the same approach for the velocity,
but we will assume here that the mean velocity $\bbV=0$.

We first consider  $\bbB=0$ for which 
equation for the total fluctuating magnetic field 
(written in  Alfv\'en speed units), 
 obtained by subtracting
the average of  (\ref{0}) from  (\ref{0}), is  
\beq
\partial_t {\bfb}=
\curl(\bfv\ts \bfb)
-\curl\lb\bfv\ts \bfb\rb + \lambda\nabla^2 \bfb.
\label{1}
\ee
Dotting this
equation with $\bfb$ and averaging, we obtain
\beq
{1\over 2}\partial_t \lb b^2 \rb=
\lb\bfb\cdot\curl(\bfv\ts \bfb)\rb
+ \lambda \lb\bfb\cdot\nabla^2 \bfb\rb.
\label{2}
\ee
The effective magnetic dissipation number  $R_{M,e}$ represents an estimate of the ratio of magnitudes of  the 
second to third term in (\ref{2}).
%

When an initially laminar system is subject to 
a source of turbulence in the velocity field, 
 both the velocity and magnetic  field quickly evolve nonlinearly
to acquire turbulent spectra. Even if the magnetic initial field were weak,
 the system typically evolves to a state of near equipartition
between total kinetic and magnetic energies, but each with distinct 
spectra (\cite{schek02}, \cite{schek04}, \cite{hbd03}, \cite{hb04}).  The triple fluctuation terms
in e.g (\ref{2})  mediate the nonlinear sustenance of the 
turbulent spectra. To obtain quantitative estimates of
(\ref{4dd}) and (\ref{5dd}), we write the magnitude of the 
magnetic field and velocity 
in terms of angle-integrated kinetic and magnetic energy spectra $E_M(k)$ and $E_V(k)$ 
respectively. We have

\beq
v(k) = (2k E_V(k))^{1/2}= v_0 \left({k\over k_{0,v}}\right)^{1-q\over 2}
\ee
and
\beq
b(k) = (2k E_M(k))^{1/2}= b_0 \left({k\over k_{0,b}}\right)^{1-m\over 2},
\ee
where $q$ and $m$ are the kinetic and magnetic energy spectral indices,  
$v_0$ and $b_0$ are the magnitude of the velocity and magnetic
field on the outer turbulent scale, and $k_{0,v}, k_{0,b}$ are
 the wave numbers for the outer scales of the
assumed kinetic and magnetic spectra. 

We also allow for different inner scales, or equivalently, allow
the magnetic Prandtl number $Pr_M$ to be a free parameter.
To accommodate this, note that 
at the scale $k_\lambda^{-1}$,  the magnetic spectrum is truncated by dissipation.
We assume that the kinetic energy cascade is mediated
by local interactions so that  $\nu\sim v_\nu/k_\nu$,
where $v_\nu$ and $k_\nu$ are the velocity and wave 
number of the viscous scale.
There must also be a characteristic speed $v_{eff}$ such that
$\lambda = v_{eff}/k_\lambda$. However, the value $v_{eff}$ need
not be $v_\lambda$ (the value of $v$ at the scale $k_\lambda^{-1}$), and 
cannot be for $Pr_M>>1$, since then $v_\lambda\sim 0$.  
Larger scale velocities  fold  the field 
which can  show up as magnetic power on small scales (\cite{schek02,schek04,hbd03,hb04}).
We then  posit that $b_\lambda \sim v_{eff}$,
since $b_\lambda$ (the field at $k_\lambda$) draws its energy from $v_{eff}$.
Using the definition of $Pr_M\equiv{\nu\over \lambda}$, we then have 
\beq
\lambda\simeq 
{b_\lambda\over k_\lambda}\simeq
{1\over Pr_M}{v_\nu \over k_\nu}
\label{9}
\ee
Eq. (\ref{9}) implies 
\beq
R_M\equiv {v_0\over k_{0,v}\lambda}=
Pr_M \left({k_\nu\over k_{0,b}}\right)^{q+1\over 2}
\left({k_{0,b}\over k_{0,v}}\right)^{q+1\over 2}={v_0\over b_0}
\left({k_\lambda\over k_{0,b}}\right)^{m+1\over 2}.
\label{10}
\ee

Assuming that the  magnetic and kinetic spectra for $k \le Min[k_\lambda,k_\nu]$ are 
mediated primarily by local interactions, we can approximate 
each nonlinear term on the right of  (\ref{2}) by an integral over
the magnitude of  $k$.
For $k_{0,b}\ge k_{0,v}$, the second term of (\ref{2}) gives
\beq
|\lb\bfb\cdot\curl(\bfv\ts \bfb)\rb|\lsim{ v_0 b_0^2} 
\int_{k_{0,b}}^{k_d} 
\left({k\over k_{0,v}}\right)^{1-q\over 2}
\left({k\over k_{0,b}}\right)^{1-m} dk
= {k_{0,b} v_0 b_0^2} \left({k_{0,b}\over k_{0,v}}\right)^{1-q\over 2}\int_{1}^{k_d/k_{0,b}}
{\kappa}^{3-q-2m \over 2}\ d\kappa,
\label{4dd}
\ee
where $k_d/k_{0,b}\equiv Min [k_\nu/k_{0,b},k_\lambda/k_{0,b}]
=Min\left[{k_{0,v}\over k_{0,b}}\left({R_M \over Pr_M}\right)^{2\over q+1},\left(
{R_M b_0\over v_0}\right)^{2\over m+1}\right]$, 
using (\ref{10}).  The inequality in (\ref{4dd})
arises because
the left side includes the competing effects of 
 turbulent amplification and turbulent
diffusion of the magnetic field.
The norm of the third term of (\ref{2}), similarly averaged,  is given by 
\beq
|\lambda\lb \bfb  \nabla^2 \bfb\rb| \sim Pr_M^{-1}{v_\nu\over k_\nu}
{b_0^2} \int_{k_{0,b}}^{k_\lambda}
k \left(k\over k_{0,b}\right)^{1-m}dk= 
{k_{0,b} 
v_0b_0^2\over R_M}\left({k_{0,b}\over k_{0,v}}\right) 
\int_{1}^{k_\lambda/k_{0,b}}
\kappa^{2-m}d\kappa
\label{5dd},
\ee
where we have again used  (\ref{10}).
Because (\ref{5dd}) involves only dissipation, whereas the left side  of (\ref{4dd})
could involve amplification or diffusion,  $R_{M,e}$, estimated
by the ratio of the right side of (\ref{4dd}) to the
right side of (\ref{5dd}), represents an upper limit to the 
ratio of the left hand sides of those equations. 

To use the ratio of (\ref{4dd}) to (\ref{5dd}) to estimate $\Q$, 
we need values for  $q$, $m$, $b_0^2/v_0^2$, $R_M$ and $Pr_M$.
We will assume that $q=5/3$ and 
plot $R_{M,e}$ as a function of independent variables 
$m$ and $Pr_M$ for various values of $R_M$.
For $m> 1$, $b_0^2/v_0^2$ can be approximated
by the ratio of total magnetic to kinetic energy
because 
\beq
{b_0^2\over v_0^2}= {E_M\over E_V}{\int^{k_\nu\over k_{0,v}}_1\kappa^{-5/3}d\kappa \over\int^{k_\lambda\over k_{0,b}}_1\kappa^{-m}d\kappa }
=
{E_M\over E_V}{\int^{k_\nu\over k_{0,v}}_1\kappa^{-5/3}d\kappa \over\int^{k_\lambda\over k_{0,b}}_1\kappa^{-m}d\kappa }
={3E_M\over 2E_V}{1-(R_M/Pr_M)^{-1/2}
\over {1\over m-1}\left(1-(R_Mb_0/v_0)^{2-2m\over 1+m}\right)}
\sim {E_M\over E_V},
\label{bv}
\ee
where the similarity follows self-consistently for 
  $R_Mb_0/v_0, R_M/Pr_M>>1$. 
For modest $R_M$  and $Pr_M$, the same approximation is also satisfactory for $m=1$, as the integrals in the second term of (\ref{bv}) become ratios of logarithms. In general, $b_0^2$ is of order $v_0^2$ for the range of $m$ considered. 
Remember that $b_0$ and $v_0$ are defined  at $k_{0,b}$ and $k_{0,v}$ respectively
and the latter two wave numbers need not be equal:
Although we assume $m$ and $q$ are 
 constant over the ranges of $k_{b,0}\le k \le k_\lambda$ and
$k_{v,0}\le k \le k_\nu$ respectively, the allowance
of $k_{v,0}\ne k_{b,0}$ and $k_\lambda\ne k_\nu$ provides
flexibility in incorporating a range of spectra compatible
with MHD simulations 
(\cite{schek02}, \cite{schek04}, \cite{hbd03}, \cite{hb04}).


Figs. 1  and 2 show $R_{M,e}$ 
calculated from the ratio of (\ref{4dd}) to (\ref{5dd})
for  
$10^{-2} \le Pr_M \le {R_M\over 10}$, and $1 \le m\le 1.7$ using 
two values of $k_{v,0}/k_{b,0}$ and two values of $(b_0/v_0)^2$.
For Fig. 1, $R_M=10^3$ and for Fig. 2 $R_M=10^6$.
To ensure $k_\nu>>k_{v,0}$, we must have $R_M/Pr_M >> 1$, 
which explains why we have chosen 
the upper limits on $Pr_M$ to be $R_M/10$ in each plot.

The case of $(b_0/v_0)^2=1/4$ used in Fig. 1bd and 2bd is motivated by 
saturated  states of non-helical
MHD turbulence which show that in cases when the magnetic energy is
initially weak, it builds up to a fraction of equipartition of the total
kinetic energy when the latter is steadily driven 
(\cite{schek04}, \cite{hb04}, \cite{hbd03}, \cite{mcm04}, \cite{hb04}).
We have not required a steady state, 
but in  that case 
the two terms on the right of (\ref{1}) would exactly balance. This
could be used to set the value of $(b_0/v_0)^2$ for a given $m$, 
$k_{v,0}/k_{b,0}$, 
$Pr_M$, and $R_M$, thereby enforcing $\Q= 1$.
However, the scaling approximation for the first term on the right
of (\ref{4dd}) is crude, and does not take into account that 
terms within that term can have different signs, reducing its value
and that of $\Q$. In the steady state, 
$\Q$,  could therefore modestly exceed unity using 
our estimation procedure. As seen in Fig. 1  
however, the important point is that $R_{M,e}<<R_M$ robustly. 

The dependencies of $R_{M,e}$ on $m$, $R_M$, $Pr_M$, $k_{v,0}/k_{b,0}$,
 and   $(b_0/v_0)^2$   are evident in Figs. 1 and 2.
In each  panel,  as  $m$ increases for a fixed $R_M$, 
the magnetic spectrum steepens
and becomes more heavily weighted by low $k$ for the same value
of $(b_0/v_0)^2$, increasing $R_{M,e}$. 
A larger value of $(b_0/v_0)^2$ lowers $R_{M,e}$,
as can be seen by comparing the two rows of Fig. 1. This is expected: 
for a given $m$, increasing $(b_0/v_0)^2$ 
increases the upper bound in (\ref{5dd}) and that integral
is more sensitive to the upper bound than the integral of (\ref{4dd}).
Since $R_{M,e}\propto \left(k_{v,0}/k_{b,0}\right)^{1+q\over 2}$,
increasing $k_{v,0}/k_{b,0}$ also increases $R_{M,e}$ as evidenced by the figures.

Each panel of Fig. 1 shows that  for a fixed $m$, increasing
$Pr_M$  lowers $\Q$. 
This is because more spectral range is available for the magnetic field
at large $k$ as $Pr_M$ is increased, thereby increasing the relative weight
of high $k$ contributions of the field  in the dissipation term.
At large $Pr_M$ and low $m$,  
$\Q\rightarrow 0$ since the shallow spectrum again increases 
the relative importance of  high $k$ 
for the dissipation
term; the advection term is ruled by the 
$k_\nu<<k_\lambda$ velocity cutoff and is less sensitive to increases
in $Pr_M$ for $Pr_M >>1$.

Comparing  Figs. 1 and 2, we see that increasing  $R_M$  increases, $\Q$ 
 only  weakly, highlighting  the role of the spectra
and  the universality of $\Q << R_{M}$.  
We reiterate  that $R_{M,e}$ is  calculated from the ratio of the
advection to microphysical dissipation terms, NOT 
by approximating  any of the advection terms via a turbulent diffusivity as is commonly done
in mean field formalism  (\cite{moffatt,zeldovich}).
We discuss the latter  in  the next section.

\section{Effective dissipation number for the mean  field equation}

When the dynamical equations of a turbulent system
such as an accretion disk (\cite{ss73}, \cite{kfr}, \cite{bh98}) 
are assumed to obey a symmetry such as axisymmetry, these equations 
describe the evolution of  mean quantities, not total quantities.
Turbulence violates the global symmetry locally and only the mean quantities
obey the symmetry.
Accretion disk theory that invokes
a turbulent viscosity (\cite{ss73})
in the axisymmetric velocity equation is in fact a
mean field theory, even when not explicitly presented as such.
The same is true for the mean magnetic field
equation (\cite{moffatt,zeldovich}).
The associated turbulent diffusivities are formally constructed from  closures
that replace  mean nonlinear correlations of macroscopic turbulent fluctuations 
 in the momentum and induction equations
by   diffusion terms, 
NOT from the  replacement of microphysical dissipation terms.
This is an important distinction from the previous section which involved
no closures  on the second term of (\ref{2}); there we simply took the ratio
of that term to the third term to define $R_{M,e}$.

More explicitly, consider the equation for evolution of the 
mean magnetic field.
This equation is obtained by 
averaging (\ref{0}), which  gives (\cite{moffatt}, \cite{p79})
\beq
\partial_t\bbB = \curl \lb \bfv \ts \bfb \rb + \curl (\bbV\ts \bbB) - \curl \lambda \nabla \bbB.
\label{mf}
\ee
When the turbulence is assumed to be isotropic, three dimensional, 
and incompressible, and $\lb\bfv\ts\bfb\rb$ is time independent, 
mean field theory with the minimal $\tau$ closure 
(\cite{bf02})  gives 
\beq
\lb\bfv\ts\bfb \rb \simeq  \tau {\tilde \alpha} \bbB 
-{\tau\tilde \beta} \curl \bbB + f(\bbV),
\label{emf}
\ee
where $\tau\sim {1\over v_0k_{0,v}}$ 
is a non-linear turbulent damping time, and 
in 3-D,
$\tau \tilde \beta \lsim \lb v^2\tau \rb$ acts as a diffusion
coefficient. The $\tilde \alpha$ term, important for the helical
dynamo (\cite{moffatt}, \cite{p79}, \cite{bs05})
 survives only when conditions (such as rotation + stratification) are able to 
sustain a mean pseudoscalar. 
The last term in (\ref{emf}) schematically indicates additional 
terms that are a function
of $\bbV$.

Here we do not discuss the 
precise value of $\tilde \alpha$ and $\tilde \beta$ in (\ref{emf}), 
rather, we simply want to emphasize that these terms 
come from turbulence and would be absent without it.
In the simplest case for which the second and fourth terms of (\ref{emf})
vanish but the third term remains, we can define the  effective magnetic Reynolds number 
for the mean field equation (\ref{mf}) as 
 ${\tilde R}_{M,e}={|\curl(\bbV\ts\bbB)|\over |\curl (\tau \tilde \beta+\lambda)(\curl \bbB)
|}$.
In terms of the velocity spectrum, if $\lambda << \tau \tilde \beta$, 
we can approximate the total diffusion by the latter and bound it by its 
upper limit such that 
\beq
\tau\tilde\beta +\lambda \sim \ta\tilde\beta \lsim \lb v^2\tau \rb\sim {v_0\over k_{0,v} }\int_{1}^{k_\nu/k_{0,v}} \kappa^{-(q+3)/2}d\kappa.
\ee
For $q=5/3$, $\tau\tilde \beta \sim v_0/k_{0,v}$. This 
highlights that  in the presence of turbulence, 
the mean field can  diffuse 
at a rate determined by a macroscopic
turbulent diffusion coefficient rather than 
the microphysical diffusion coefficient. 
That is, 
\beq
{\tilde R}_{M,e} \le   {\bV L\over \tau \tilde \beta} = {\bV \over  L k_{0,v} v_0},
\label{cont}
\ee
which can be unity for $\bV\sim v_0$ and $Lk_{0,v}\sim 1$.

{We emphasize again that $R_{M,e}$ of the previous section has a different
definition from  ${\tilde R}_{M,e}$ just defined.
The former measures the ratio of advection terms to the microphysical dissipation terms in the total energy equation, 
while the latter applies only to the mean field equation and 
depends on our having introduced a macroscopic turbulent diffusivity $\tau \tilde \beta$ to replace some triple fluctuation terms within the advection term by a
diffusion term. 
The total mean field diffusion coefficient is sum of
turbulent and microphysical diffusivities and so 
the denominator of ${\tilde R}_{M,e}$ 
also includes contributions from both.
When the turbulent  diffusivity  greatly exceeds the microphysical diffusivity, the the denominator of ${\tilde R}_{M,e}$ 
is dominated by the former. This contrasts the denominator of 
${R}_{M,e}$ which  only involves microphysical dissipation.
The reason that ${R}_{M,e}$ can also be small, as discussed in the previous section,  is that small scale structures lead to a substantial
microphysical dissipation term even if the microphysical diffusion coefficient is small.

{That ${\tilde R}_{M,e}$ and $R_{M,e}$ can be small and independent of $R_M$
results because turbulence cascades large scale structures down to  microphysical scales
where dissipation  is fast.
For driven turbulence, the overall rate at which the cascade occurs 
is fixed by the forcing. As long as there is some microphysical mechanism
of energy drain at small scales the rate of that mechanism
does not strongly influence the overall rate of total energy dissipation; 
the system will develop the structures needed to 
accommodate the cascade rate.   
}


The extent to which 
 mean field turbulent diffusion in MHD is suppressed below the maximum
kinematic macroscopic value 
by the backreaction from fluctuating and mean magnetic fields has been
a topic of study
(\cite{vc92,c94,b96,gd95,rk01,bb02,bs05}). 
Part of the reason for some 
apparent disagreement  
is that different researchers have made different simplifying assumptions 
(boundary conditions, steady vs. evolving states, 2-D vs. 3-D)
and although each may present a result consistent with their assumptions
the universality of the conclusions still needs to be sorted out.
The suppression
seems to be less dramatic in 3-D. 
More simulations are needed and 
further detailed discussion is beyond the scope
of the present paper. To accommodate a range of probabilities
we have  written an inequality  
in (\ref{cont}), but the basic principles discussed as to why a  
cascade  would dictate   $R_{M,e}<<1$ and ${\tilde R}_{M,e} < R_M$ remain. 

As discussed in Sec 1., the principles regarding the importance of 
the dissipation terms derive from, and are analogous to
similar circumstances in the purely 
 hydrodynamic case, with analogous effective dissipation numbers defined appropriately.  In the absence of magnetic fields however, there is no backreaction so 
the hydrodynamic evolution of the mean velocity field and its turbulent diffusion 
are not subject to the issues raised in the previous paragraph.

\section{ Conclusions}


{
We have  calculated a  
dimensionless measure of the relative importance of advective vs. dissipative terms for the total
magnetic energy equation in MHD that applies for arbitrary magnetic and 
kinetic spectra, and magnetic Prandtl number.
We have discussed how 
the standard magnetic Reynolds number $R_M$ defined using the largest gradient scale
of a system does not  accurately  approximate this ratio for turbulent systems, even if the magnetic energy is dominated by large scale structures.  
Instead, this ratio is best approximated by  $R_{M,e}$, a quantity
 largely independent of $R_M$ 
for a range of turbulent spectra (though reverting to $R_M$ for a laminar flow) and
 typically satisfies $R_{M,e}\sim O(1) <<R_M$.
The latter results because 
 small scale structures augment the relative importance of the 
 microphysical dissipation terms for a turbulent system 
 compared to a laminar system with the same microphysical diffusion coefficient.
In a steady state,  such as those  reached when MHD 
turbulence is steadily forced, 
$R_{M,e}\sim 1$.

We have also discussed how the effective dissipation number ${\tilde R}_{M,e}$ 
for the mean magnetic field equation is defined differently
from $R_{M,e}$ because 
the mean field equation picks out the gradient scale of the
mean field only.  A low value of ${\tilde R}_{M,e}<<R_M$
 results not from small scale gradients, but
 because macroscopic turbulent diffusion coefficients exceed  microscopic
diffusivities in closures that approximate  nonlinear 
cascade terms by diffusion.

While the results that ${\tilde R}_{M,e}<<R_M$, $ R_{M,e}<<R_M$, 
are  not at all surprising when interpreting them analogously to
steady-state Kolmogorov hydrodynamic turbulence, they 
do highlight the much greater role of dissipation 
in large $R_M$ turbulent  MHD flows compared to large $R_M$ laminar  flows.
In particular, an important implication  of  ${\tilde R}_{M,e}<<R_M$, $ R_{M,e}<<R_M$, 
and the very weak dependence of  $R_{M,e}$ and ${\tilde R}_{M,e}$ on
 $R_M$  is that the rate at which large scale MHD energy is dissipated
can be much less sensitive to the actual microphysical dissipation mechanism
in turbulent flows when compared to laminar flows.
On a macroscopic dynamical time, the MHD energy
cascades to scales small enough such that  it can be drained at multiple sites. If the turbulence is steadily forced, then the 
system adjusts to produce enough small scale structures to accommodate the driving
energy input rate.  The overall energy dissipation rate 
can  be insensitive to the microphysical reconnection  (or other microphysical energy conversion rate) at any one site. Instead, the aggregate of dissipation sites emerges 
 to drain the total magnetic energy  on macroscopic dynamical 
 time scales. Since many astrophysical settings are MHD turbulent,
this highlights the importance of being precise when asking and studying the question ``Is magnetic reconnection fast in astrophysics?''

The principle that small scale structures increase the effective reconnection rate is consistent with  rapid reconnection 
mechanisms in the MHD  models of Hendrix et al. (1996), Galasgaard \& Nordlund (1996), Lazarian \& Vishniac (1999), 
and in the rapid reconnection phase of  Fan et al. (2004).
Reconnection studies that start with a large scale laminar field
and reconnect on collisionless scales (e.g. \cite{birn,shay})
 address a different regime.  The latter can however apply to each of the small scale 
 multiple dissipation sites within a globally turbulent MHD system.
In some systems, collisionless individual structures are observationally resolved (e.g. the Earth's magnetosphere) and studying the plasma instabilities and plasma turbulence for different conditions on such scales  
is needed to understand when reconnecting structures  incur a 
reconnection  rate that depends only on its macroscopic properties (\cite{kulsrud}).
That being said, if a stellar or accretion disk corona  is considered as a 
global entity, its overall energy dissipation rate in a steady state is determined by the dynamical rate of 
magnetohydrodynamic energy injection from below; the small scale structures develop to accommodate this rate.  The 
overall dissipation rate is best characterized by ${\tilde R}_M<<R_M$ and $R_{M,e}<<R_M$ even if the rate of each individual dissipation site were to depend on  $R_M$.



%

\ni {\bf Acknowledgments:} 
EGB acknowledges support from 
NSF grant AST-0406799 and NASA grant ATP04-0000-0016.

\eject

\vspace{-.1cm} \hbox to \hsize{ \hfill 
\hspace{-1cm}
\epsfxsize8cm
\epsffile{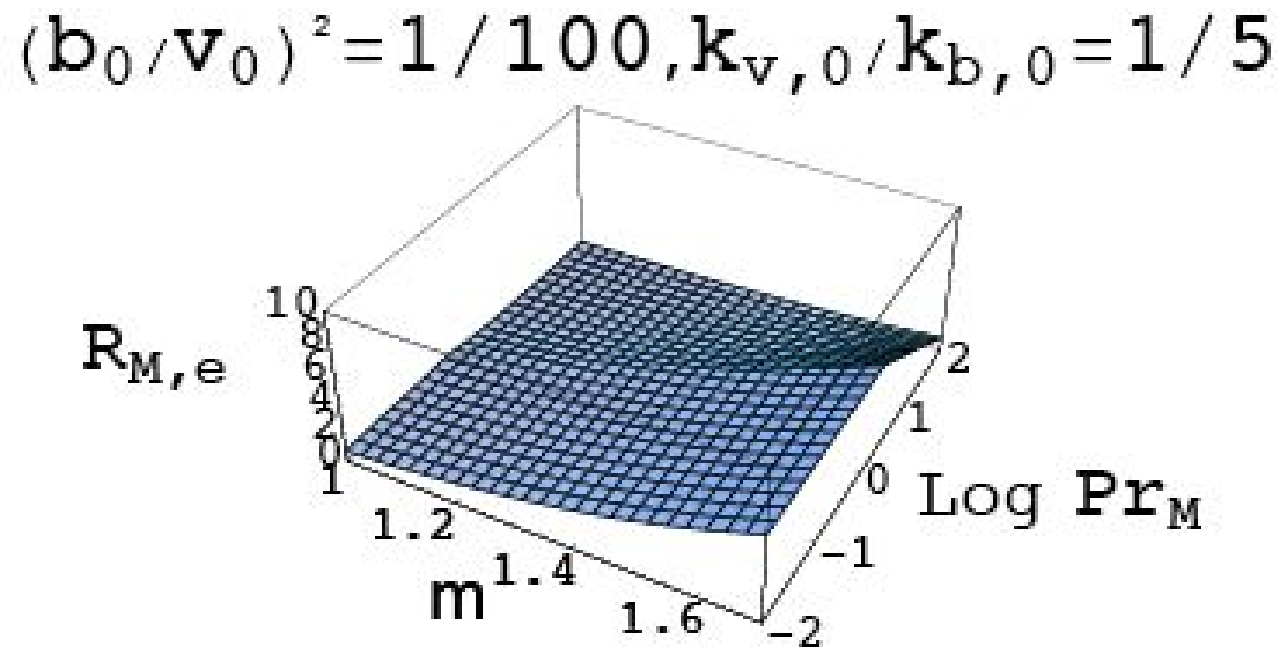} \hspace{0.7cm}
\epsfxsize8cm
\epsffile{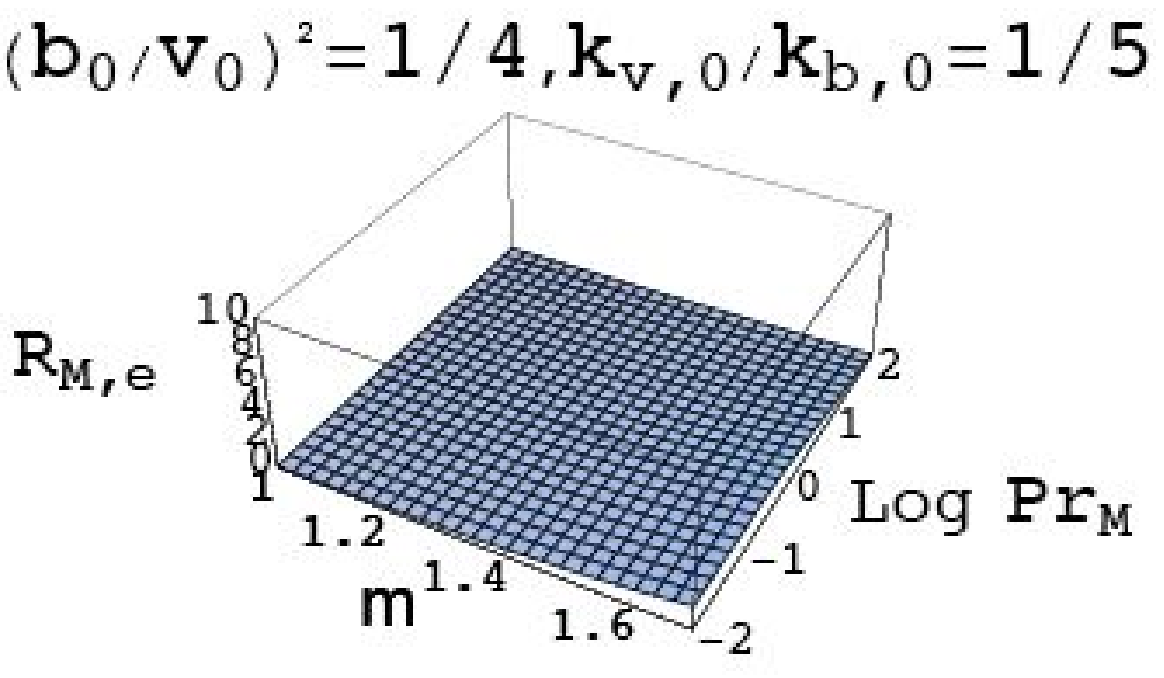} 
\hfill }
\hspace{-1cm} \hbox to \hsize{ \hfill \epsfxsize8cm 
\epsffile{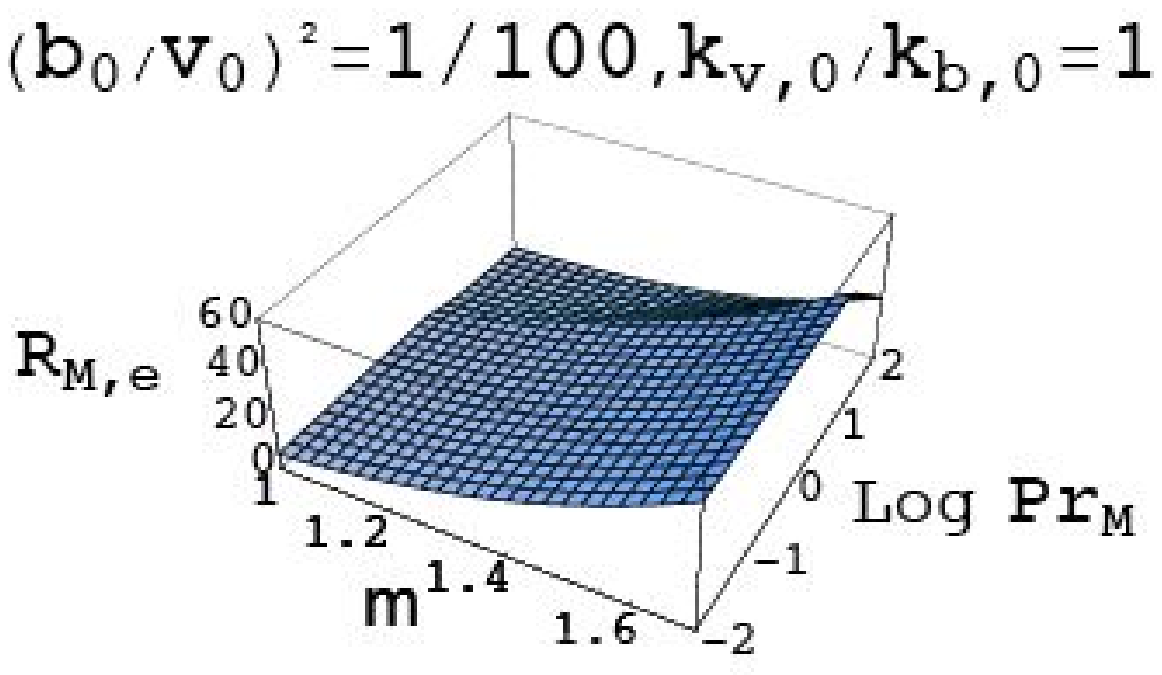} \hspace{0.7cm}\epsfxsize8cm \epsffile{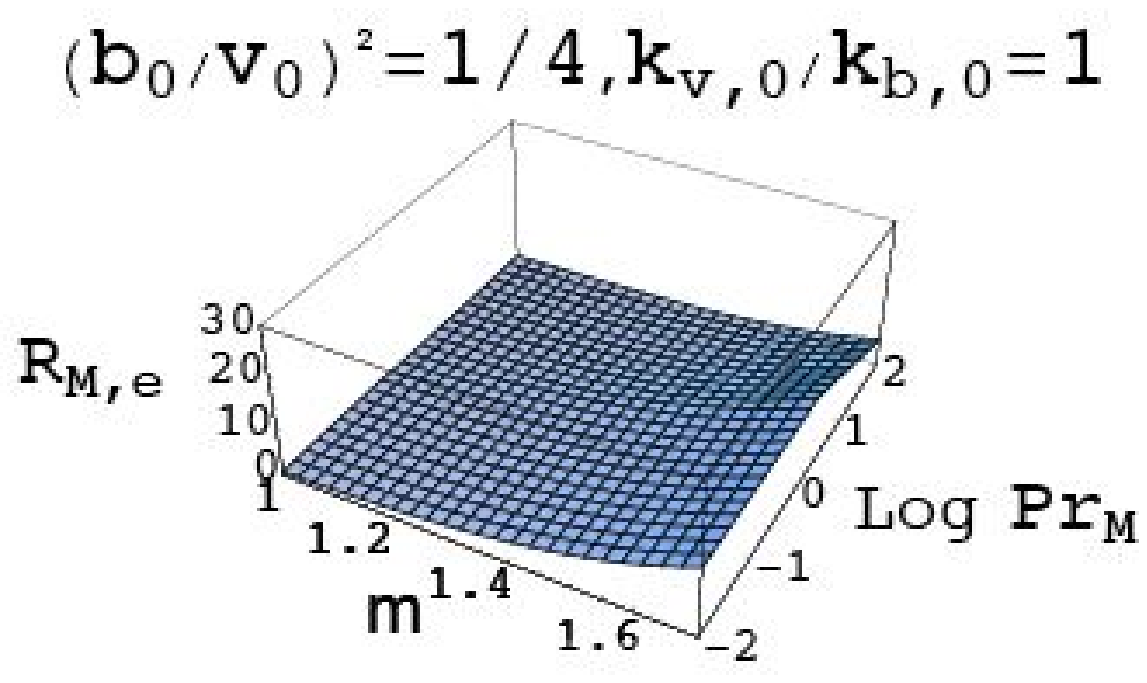} 
\hfill }

\ni {Figure 1:  $R_{M,e}$  vs. 
 magnetic spectral index $m$ and magnetic Prandtl number
$Pr_M$ for $R_M=10^3$. 
a) $(b_0/v_0)^2=1/100$, $k_{v,0}/k_{b,0}=1/5$
b) $(b_0/v_0)^2=1/4$, $k_{v,0}/k_{b,0}=1/5$
c) $(b_0/v_0)^2=1/100$, $k_{v,0}/k_{b,0}=1$
d) $(b_0/v_0)^2=1/4$, $k_{v,0}/k_{b,0}=1$
Note the weak dependence of $R_{M,e}$ on $R_M$.}

\eject
\vspace{-.1cm} \hbox to \hsize{ \hfill 
\hspace{-1cm}
\epsfxsize8cm
\epsffile{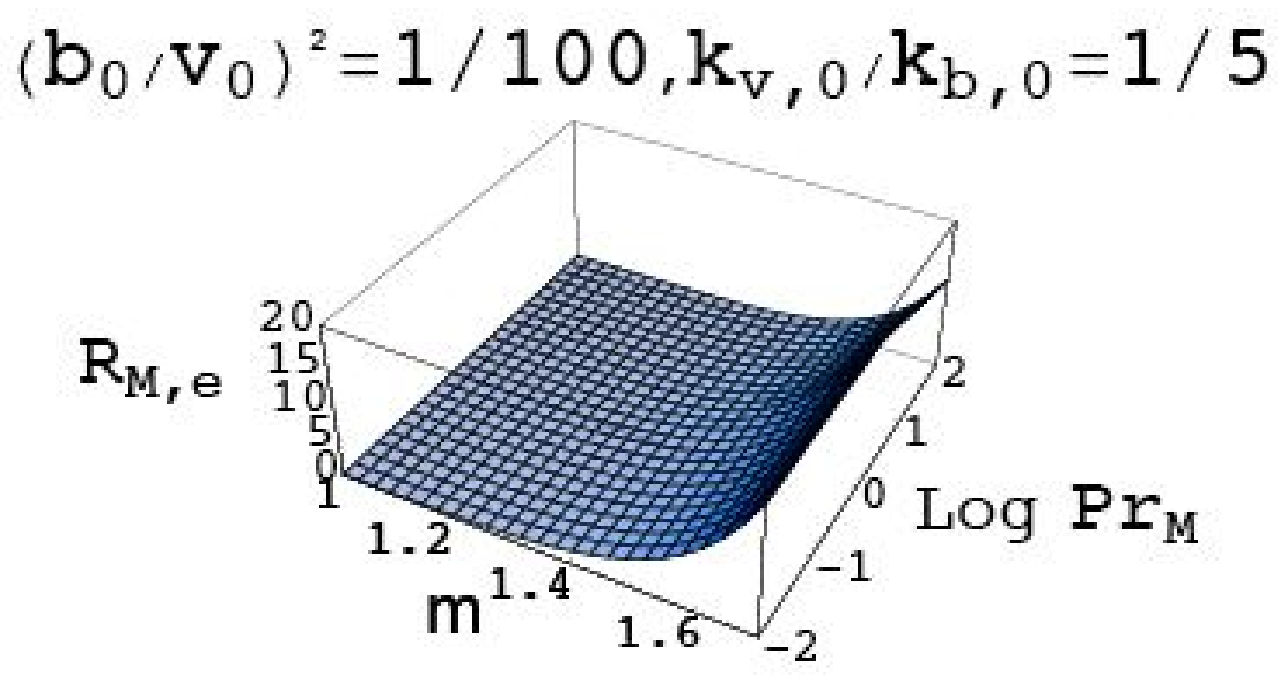} \hspace{0.5in}
\epsfxsize8cm
\epsffile{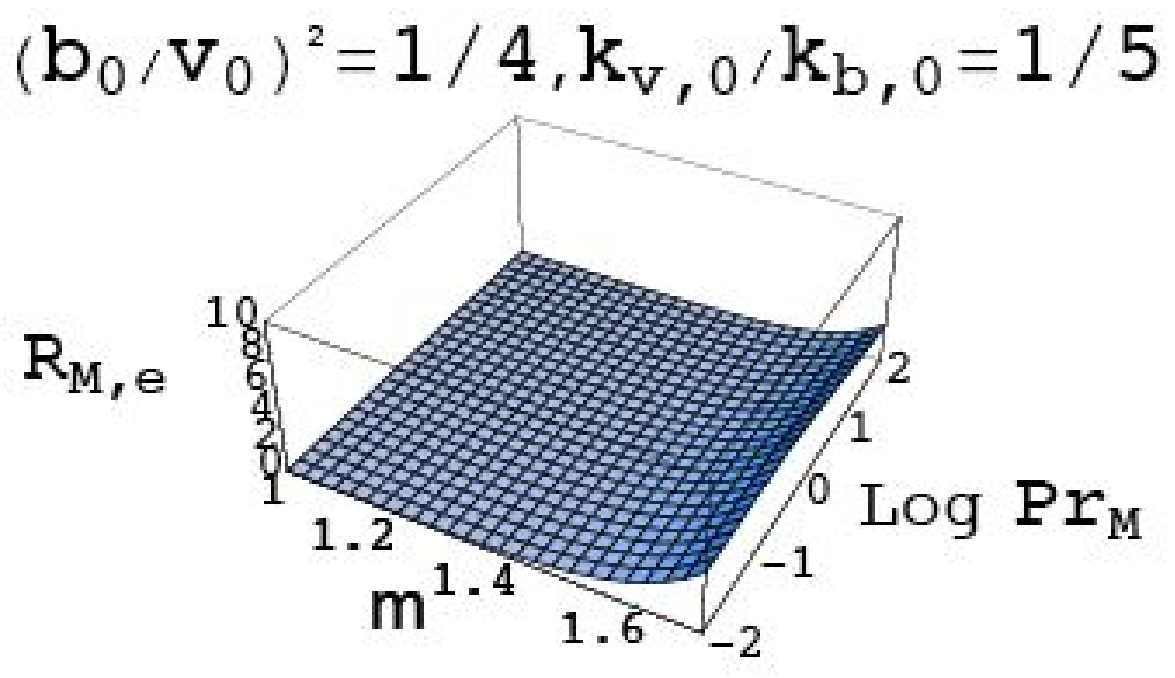} 
\hfill }
\hspace{-1cm} \hbox to \hsize{ \hfill \epsfxsize8cm 
\epsffile{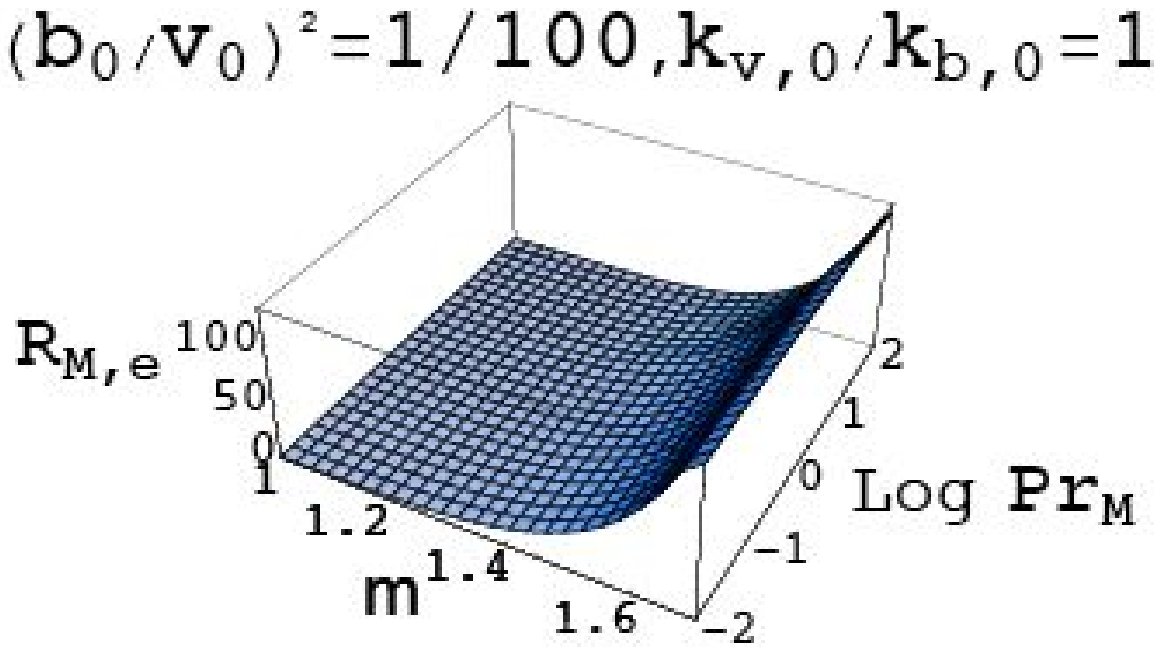} \hspace{0.5in}\epsfxsize8cm \epsffile{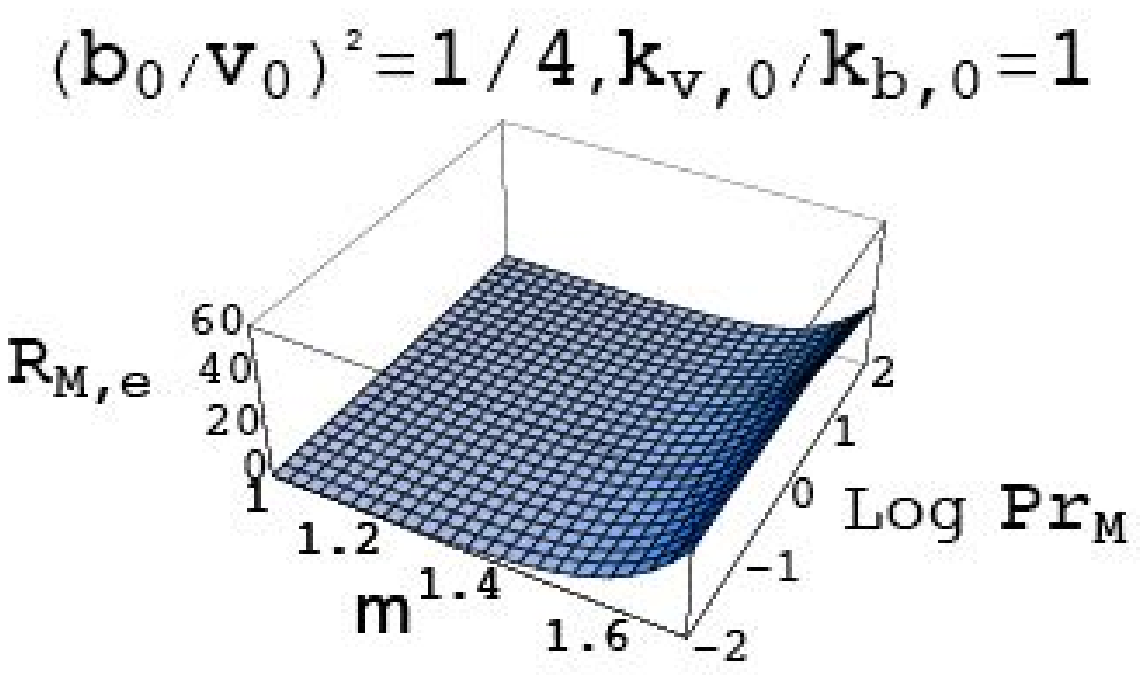} 
\hfill }

\ni {Figure 2:  
Same as Fig. 1 but with $R_M=10^6$}
\eject


\end{document}